\documentclass[
  journal=pasa,
  manuscript=research-paper, 
  year=2023,
]{cup-journal}

\usepackage{microtype,siunitx,booktabs}
\newcommand{\colhead}[1]{}

\newcommand\monash{School of Physics and Astronomy, Monash University, VIC 3800, Australia}

\RequirePackage[normalem]{ulem} 
\RequirePackage{color}\definecolor{RED}{rgb}{1,0,0}\definecolor{BLUE}{rgb}{0,0,1} 

\title{The Parkes Pulsar Timing Array Third Data Release}

\author{Andrew Zic}
\affiliation{Australia Telescope National Facility, CSIRO, Space and Astronomy, PO Box 76, Epping, NSW 1710, Australia}
\alsoaffiliation{Department of Physics and Astronomy and MQ Research Centre in Astronomy, Astrophysics and Astrophotonics, Macquarie University, NSW 2109, Australia}
\email[A. Zic, D. J. Reardon]{andrew.zic@csiro.au, dreardon@swin.edu.au}

\author{Daniel J. Reardon}
\affiliation{Centre for Astrophysics and Supercomputing, Swinburne University of Technology, P.O. Box 218, Hawthorn, Victoria 3122, Australia}
\alsoaffiliation{OzGrav: The Australian Research Council Centre of Excellence for Gravitational Wave Discovery, Hawthorn VIC 3122, Australia}

\author{Agastya Kapur}
\affiliation{Australia Telescope National Facility, CSIRO, Space and Astronomy, PO Box 76, Epping, NSW 1710, Australia}
\alsoaffiliation{Department of Physics and Astronomy and MQ Research Centre in Astronomy, Astrophysics and Astrophotonics, Macquarie University, NSW 2109, Australia}

\author{George Hobbs}
\affiliation{Australia Telescope National Facility, CSIRO, Space and Astronomy, PO Box 76, Epping, NSW 1710, Australia}

\author{Rami Mandow}
\affiliation{Department of Physics and Astronomy and MQ Research Centre in Astronomy, Astrophysics and Astrophotonics, Macquarie University, NSW 2109, Australia}
\alsoaffiliation{Australia Telescope National Facility, CSIRO, Space and Astronomy, PO Box 76, Epping, NSW 1710, Australia}

\author{Ma\l{}gorzata Cury\l{}o}
\affiliation{Astronomical Observatory, University of Warsaw, Aleje Ujazdowskie 4, 00-478 Warsaw, Poland}

\author{Ryan M. Shannon}
\affiliation{Centre for Astrophysics and Supercomputing, Swinburne University of Technology, P.O. Box 218, Hawthorn, Victoria 3122, Australia}
\alsoaffiliation{OzGrav: The Australian Research Council Centre of Excellence for Gravitational Wave Discovery, Hawthorn VIC 3122, Australia}

\author{Jacob Askew}
\affiliation{Centre for Astrophysics and Supercomputing, Swinburne University of Technology, P.O. Box 218, Hawthorn, Victoria 3122, Australia}
\alsoaffiliation{OzGrav: The Australian Research Council Centre of Excellence for Gravitational Wave Discovery, Hawthorn VIC 3122, Australia}

\author{Matthew Bailes}
\affiliation{Centre for Astrophysics and Supercomputing, Swinburne University of Technology, P.O. Box 218, Hawthorn, Victoria 3122, Australia}
\alsoaffiliation{OzGrav: The Australian Research Council Centre of Excellence for Gravitational Wave Discovery, Hawthorn VIC 3122, Australia}

\author{N.~D.~Ramesh Bhat}
\affiliation{International Centre for Radio Astronomy Research, Curtin University, Bentley, WA 6102, Australia}

\author{Andrew Cameron}
\affiliation{Centre for Astrophysics and Supercomputing, Swinburne University of Technology, P.O. Box 218, Hawthorn, Victoria 3122, Australia}
\alsoaffiliation{OzGrav: The Australian Research Council Centre of Excellence for Gravitational Wave Discovery, Hawthorn VIC 3122, Australia}

\author{Zu-Cheng Chen}
\affiliation{Advanced Institute of Natural Sciences, Beijing Normal University, Zhuhai 519087, China}
\alsoaffiliation{Department of Physics and Synergistic Innovation Center for Quantum Effects and Applications, Hunan Normal University, Changsha, Hunan 410081, China}

\author{Shi Dai}
\affiliation{School of Science, Western Sydney University, Locked Bag 1797, Penrith South DC, NSW 2751, Australia}

\author{Valentina Di Marco}
\affiliation{\monash}
\alsoaffiliation{OzGrav: The Australian Research Council Centre of Excellence for Gravitational Wave Discovery, Clayton VIC 3800, Australia}

\author{Yi Feng}
\affiliation{Research Center for Intelligent Computing Platforms, Zhejiang Laboratory, Hangzhou 311100, China}

\author{Matthew Kerr}
\affiliation{Space Science Division, US Naval Research Laboratory, 4555 Overlook Ave SW, Washington DC 20375, USA}

\author{Atharva Kulkarni}
\affiliation{Centre for Astrophysics and Supercomputing, Swinburne University of Technology, P.O. Box 218, Hawthorn, Victoria 3122, Australia}
\alsoaffiliation{OzGrav: The Australian Research Council Centre of Excellence for Gravitational Wave Discovery, Hawthorn VIC 3122, Australia}

\author{Marcus E. Lower}
\affiliation{Australia Telescope National Facility, CSIRO, Space and Astronomy, PO Box 76, Epping, NSW 1710, Australia}

\author{Rui Luo}
\affiliation{Department of Astronomy, School of Physics and Materials Science, Guangzhou University, Guangzhou 510006, China}
\alsoaffiliation{Australia Telescope National Facility, CSIRO, Space and Astronomy, PO Box 76, Epping, NSW 1710, Australia}

\author{Richard N. Manchester}
\affiliation{Australia Telescope National Facility, CSIRO, Space and Astronomy, PO Box 76, Epping, NSW 1710, Australia}

\author{Matthew T. Miles}
\affiliation{Centre for Astrophysics and Supercomputing, Swinburne University of Technology, P.O. Box 218, Hawthorn, Victoria 3122, Australia}
\alsoaffiliation{OzGrav: The Australian Research Council Centre of Excellence for Gravitational Wave Discovery, Hawthorn VIC 3122, Australia}

\author{Rowina S. Nathan}
\affiliation{\monash}
\alsoaffiliation{OzGrav: The Australian Research Council Centre of Excellence for Gravitational Wave Discovery, Clayton VIC 3800, Australia}

\author{Stefan Os{\l}owski}
\affiliation{Manly Astrophysics, 15/41-42 East Esplanade, Manly, NSW 2095, Australia}

\author{Axl F. Rogers}
\affiliation{Institute for Radio Astronomy \& Space Research, Auckland University of Technology, Private Bag 92006, Auckland 1142, New Zealand}

\author{Christopher J. Russell}
\affiliation{CSIRO Scientific Computing, Australian Technology Park, Locked Bag 9013, Alexandria, NSW 1435, Australia}

\author{John~M.~Sarkissian}
\affiliation{Australia Telescope National Facility, CSIRO, Space and
Astronomy, Parkes Observatory, PO Box 276, Parkes, NSW, 2870, Australia}

\author{Mohsen Shamohammadi}
\affiliation{Centre for Astrophysics and Supercomputing, Swinburne University of Technology, P.O. Box 218, Hawthorn, Victoria 3122, Australia}
\alsoaffiliation{OzGrav: The Australian Research Council Centre of Excellence for Gravitational Wave Discovery, Hawthorn VIC 3122, Australia}

\author{Ren\'ee Spiewak}
\affiliation{Jodrell Bank Centre for Astrophysics, Department of Physics and Astronomy, University of Manchester, Manchester M13 9PL, UK}

\author{Nithyanandan Thyagarajan}
\affiliation{Australia Telescope National Facility, CSIRO, Space \& Astronomy, P. O. Box 1130, Bentley, WA 6102, Australia}

\author{Lawrence Toomey}
\affiliation{Australia Telescope National Facility, CSIRO, Space and Astronomy, PO Box 76, Epping, NSW 1710, Australia}

\author{Shuangqiang Wang}
\affiliation{Xinjiang Astronomical Observatory, Chinese Academy of Sciences, Urumqi, Xinjiang 830011, China}

\author{Lei Zhang}
\affiliation{National Astronomical Observatories, Chinese Academy of Sciences, A20 Datun Road, Chaoyang District, Beijing 100101, People's Republic of China}
\alsoaffiliation{Australia Telescope National Facility, CSIRO, Space and Astronomy, PO Box 76, Epping, NSW 1710, Australia}

\author{Songbo Zhang}
\affiliation{Purple Mountain Observatory, Chinese Academy of Sciences, Nanjing 210008, China}
\alsoaffiliation{Australia Telescope National Facility, CSIRO, Space and Astronomy, PO Box 76, Epping, NSW 1710, Australia}

\author{Xing-Jiang Zhu}
\affiliation{Advanced Institute of Natural Sciences, Beijing Normal University, Zhuhai 519087, China}

\received {08 May 2023}
\revised  {dd Mmm YYYY}
\accepted {dd Mmm YYYY}
\published{29 June 2023}


\begin{document}

\begin{abstract}
We present the third data release from the Parkes Pulsar Timing Array (PPTA) project. The release contains observations of 32 pulsars obtained using the 64-m Parkes ``Murriyang'' radio telescope. The data span is up to 18 years with a typical cadence of 3 weeks. This data release is formed by combining an updated version of our second data release with $\sim 3$ years of more recent data primarily obtained using an ultra-wide-bandwidth receiver system that operates between 704 and 4032\,MHz. We provide calibrated pulse profiles, flux-density dynamic spectra, pulse times of arrival, and initial pulsar timing models. We describe methods for processing such wide-bandwidth observations, and compare this data release with our previous release. 
\end{abstract}

\section{Introduction}
Pulsar timing arrays (PTAs) observe millisecond pulsars (MSPs) with the primary goal of detecting nanohertz-frequency gravitational waves (GWs). Current PTA collaborations include the European Pulsar Timing Array \citep[EPTA; ][]{2013CQGra..30v4009K}, the Indian Pulsar Timing Array \citep[InPTA; ][]{2018JApA...39...51J}, the North American Nanohertz Observatory for Gravitational waves (NANOGrav; \citealp{2013CQGra..30v4008M}), and the Parkes Pulsar Timing Array \citep[PPTA; ][]{2013PASA...30...17M}, all of which constitute the International Pulsar Timing Array \citep[IPTA; ][]{2013CQGra..30v4010M}.
Other collaborations, such as the Chinese Pulsar Timing Array \citep[CPTA;][]{2016ASPC..502...19L}, MeerKAT Pulsar Timing Array \cite[MPTA;][]{2023MNRAS.519.3976M}, the Fermi Pulsar Timing Array \citep{2022Sci...376..521F}, and CHIME/Pulsar \citep{2021ApJS..255....5C, Good_2021} are also constructing high-quality data sets in the effort to detect nanohertz-frequency GWs.

Beyond detection of GWs, other key science aims of PTA experiments include constraining the solar system ephemeris \citep{2010ApJ...720L.201C,2018MNRAS.481.5501C,2019MNRAS.489.5573G,2020ApJ...893..112V}, developing a pulsar-based timescale \citep{2012MNRAS.427.2780H, 2020MNRAS.491.5951H}, monitoring the interstellar medium \citep{2013MNRAS.429.2161K,2017ApJ...841..125J, 2021A&A...651A...5K, 2022PASA...39...53T}, heliosphere, and interplanetary medium \citep{2007ApJ...671..907Y, 2012MNRAS.422.1160Y, 2021A&A...647A..84T, 2022ApJ...929...39H}, and studying MSPs as individuals or as a population \citep[e.g., ][]{2011MNRAS.418.1258O, 2015MNRAS.454.1058L, 2015MNRAS.449.3223D, 2022MNRAS.510.5908M, 2022arXiv221012266J}.

The PPTA has been on-going since 2004 and uses the 64-m-diameter Parkes radio telescope, which has been gifted the Wiradjuri name ``Murriyang'', which we use hereafter. The PPTA is a major observing project at the observatory and continues to obtain approximately 48 hours of observations every 2--3 weeks. 
The PPTA, along with the EPTA and NANOGrav, have now amassed data of sufficient duration and quality that the detection of a stochastic GW background (GWB) is plausible. Data taken under the respective programs are now being analysed to search for and scrutinise any noise process that has a common spectrum across all pulsars and exhibits spatial correlations expected for the GWB 
\cite[e.g., ][]{2020ApJ...905L..34A,2021ApJ...917L..19G,2021MNRAS.508.4970C,2022ApJ...932L..22G,2022MNRAS.516..410Z}. 
The data sets from individual collaborations can be combined together, improving the sensitivity to GWs \citep{2013CQGra..30v4015S}. The IPTA data release 2 \citep{2019MNRAS.490.4666P} was formed by combining a subset of PPTA Data Release 2 \citep{2020PASA...37...20K} with the EPTA DR1 \citep{2016MNRAS.458.3341D} and the NANOGrav 9-year data set \citep{2015ApJ...813...65N}. \citet{2022MNRAS.510.4873A} presented a search for a common-spectrum process and spatial correlations expected of the GWB in the IPTA DR2.

%

\begin{table*}
\caption{Summary of the pulsars and observations forming the Parkes Pulsar Timing Array data release 3. Columns 5 through 7 give the observing time-span, number of observations, and total number of ToAs from the new UWL observations. Columns 8 through 10 are similar but for the whole DR3, and column 11 gives the legacy data span where available. Pulsars labelled with an asterisk are newly added to our data set. \label{tb:dr3}}
\centering
\begin{tabular}{lccccccccccc}
\hline
PSR & $P$ & DM & $P_b$  & T$^{\rm UWL}_{\rm span}$ & N$^{\rm UWL}_{\rm obs}$ & N$^{\rm UWL}_{\rm toas}$ & T$^{\rm DR3}_{\rm span}$ & N$^{\rm DR3}_{\rm obs}$ & N$^{\rm DR3}_{\rm toas}$ & T$^{\rm legacy}_{\rm span}$  \\
 & (ms) & (cm$^{-3}$\,pc) & (d) & (yr) &  &  & (yr) &  &  & (yr)  \\
 \hline
J0030$+$0451$^*$        & 4.87  & 4.3   & ---           &  3.17  & 36 & 593       &   3.17 & 36 & 593 & --- \\
J0125$-$2327$^*$        & 3.68  & 9.6   & 7.28          &  3.19  &  105  &  2706  &   3.19 & 105 & 2706 & --- \\
J0437$-$4715            & 5.76  & 2.6   & 5.74          &  3.80  &  216  &  4780  &   14.64 & 1854 & 11637 & 25.89 \\
J0613$-$0200            & 3.06  & 38.8  & 1.20          &  3.71  &  80  &  1777   &   18.07 & 928 & 4927 & 22.23 \\
J0614$-$3329$^*$        & 3.15  & 37.0  & 53.58         &  3.19  &  80  &  698    &   3.19 & 80 & 698 & --- \\
J0711$-$6830            & 5.49  & 18.4  & ---           &  3.71  &  170  &  2734  &   18.08 & 1011 & 5538 & 28.12 \\
J0900$-$3144$^*$        & 11.11 & 75.7  & 18.74         &  2.79  &  66  &  1832   &   2.79 & 66 & 1832 & --- \\
J1017$-$7156            & 2.34  & 94.2  & 6.51          &  3.71  &  195  &  3738  &   11.64 & 818 & 5887 & --- \\
J1022$+$1001            & 16.45 & 10.3  & 7.81          &  3.79  &  92  &  1416   &   18.08 & 983 & 5242 & 19.15 \\
J1024$-$0719            & 5.16  & 6.5   & ---           &  3.29  &  41  &  628    &   18.08 & 400 & 1841 & 26.09 \\
J1045$-$4509            & 7.47  & 58.1  & 4.08          &  3.70  &  55  &  1308   &   18.08 & 747 & 4316 & 28.04 \\
J1125$-$6014            & 2.63  & 52.9  & 8.75          &  3.66  &  97  &  2000   &   14.20 & 386 & 2832 & --- \\
J1446$-$4701            & 2.19  & 55.8  & 0.28          &  3.29  &  61  &  444    &   11.09 & 193 & 732 & --- \\
J1545$-$4550            & 3.58  & 68.4  & 6.20          &  3.71  &  100  &  1528  &   10.84 & 363 & 2454 & --- \\
J1600$-$3053            & 3.60  & 52.3  & 14.35         &  3.71  &  58  &  1303   &   18.08 & 866 & 5146 & 20.11 \\
J1603$-$7202            & 14.84 & 38.0  & 6.31          &  3.71  &  114  &  2392  &   18.08 & 821 & 5141 & 26.34 \\
J1643$-$1224            & 4.62  & 62.4  & 147.02        &  3.79  &  46  &  1156   &   18.07 & 670 & 4183 & 27.99 \\
J1713$+$0747            & 4.57  & 16.0  & 67.83         &  2.77  &  54 &  1301    &   17.15 & 886 & 5141 & 27.06 \\
J1730$-$2304            & 8.12  & 9.6   & ---           &  3.79  &  41  &  982    &   18.08 & 611 & 3306 & 27.99 \\
J1741$+$1351$^*$        & 3.75  & 24.2  & 16.34         &  2.56  &  15  & 111     &   2.56 & 15 & 111 & --- \\
J1744$-$1134            & 4.07  & 3.1   & ---           &  3.79  &  88  &  2053   &   18.08 & 884 & 5401 & 27.15 \\
J1824$-$2452A           & 3.05  & 119.9 & ---           &  2.06  &  11  &  214    &   17.02 & 317 & 1284 & --- \\
J1832$-$0836            & 2.72  & 28.2  & ---           &  2.68  &  20  &  172    &   9.16 & 88 & 385 & --- \\
J1857$+$0943            & 5.36  & 13.3  & 12.33         &  3.12  &  25  &  617    &   18.02 & 507 & 2594 & 17.90 \\
J1902$-$5105$^*$        & 1.74  & 36.2  & 2.01          &  2.85  &  38  &  501    &   2.85 & 38  & 501  & --- \\
J1909$-$3744            & 2.95  & 10.4  & 1.53          &  3.79  &  263  &  4104  &   18.08 & 1847 & 9644 & 19.24 \\
J1933$-$6211$^*$        & 3.54  & 11.5  & 12.82         &  3.17  &  83  &  893    &   3.17 & 83 & 892 & --- \\
J1939$+$2134            & 1.56  & 71.0  & ---           &  2.83  &  15  &  413    &   17.75 & 287 & 1473 & 26.31 \\
J2124$-$3358            & 4.93  & 4.6   & ---           &  3.71  &  71  &  1083   &   18.08 & 686 & 3411 & 27.81 \\
J2129$-$5721            & 3.73  & 31.9  & 6.63          &  3.70  &  114  &  1290  &   17.75 & 696 & 2921 & 26.44 \\
J2145$-$0750            & 16.05 & 9.0   & 6.84          &  3.30  &  58  &  1383   &   17.96 & 854 & 4944 & 27.73 \\
J2241$-$5236            & 2.19  & 11.4  & 0.15          &  3.70  &  164  &  3306  &      12.07 & 888 & 6238 & --- \\
\end{tabular}
\end{table*}

The second PPTA data release \citep{2020PASA...37...20K} presented our observations through early 2018, prior to the commissioning of a new receiver system. This receiver -- the ultra-wide-bandwidth low-frequency receiver \citep[UWL; ][]{2020PASA...37...12H} --  observes with continuous radio-frequency coverage between 704 and 4032\,MHz, and has been the primary receiver used for PPTA observations since late 2018. The wide bandwidth of the UWL affords a slew of benefits for the PPTA. Key among these are the increased instantaneous sensitivity. Furthermore, UWL observations are excellent for measuring variations in pulse profiles across wide bandwidths (see \citealt{2015MNRAS.449.3223D}), which has now been carefully accounted for using wide-band timing techniques \citep{2014ApJ...790...93P, 2016ascl.soft06013P, 2023ApJ...944..128C}. The large instantaneous bandwidth also enables precise instantaneous measurements of dispersion measure (DM), and therefore its long-term variations, which is a key noise process to be accounted for in GW searches \citep{2013MNRAS.429.2161K,2016ApJ...817...16C}.

In this paper, we present our latest data set (DR3), which includes new observations with the UWL (\S~\ref{sec:uwl}). We describe the data reduction pipeline used to produce our third data release, which includes a partial re-reduction of data published as part of our previous release. This resulted in an improvement to the older data. We discuss the data set in \S\ref{sec:discussion}. Our single-pulsar noise modelling and GWB analyses of this data are presented in companion papers \citep{PPTA-DR3_noise,PPTA-DR3_gwb}. The data set will also be included in the upcoming IPTA Data Release 3. The most recent data sets and GW analyses from other IPTA member collaborations are also presented in companion papers to the PPTA DR3 papers. Our data set is publicly available. Access is described in the data availability section.

\section{The data release}

\begin{figure}
    \centering
    \includegraphics[width=0.95\linewidth]{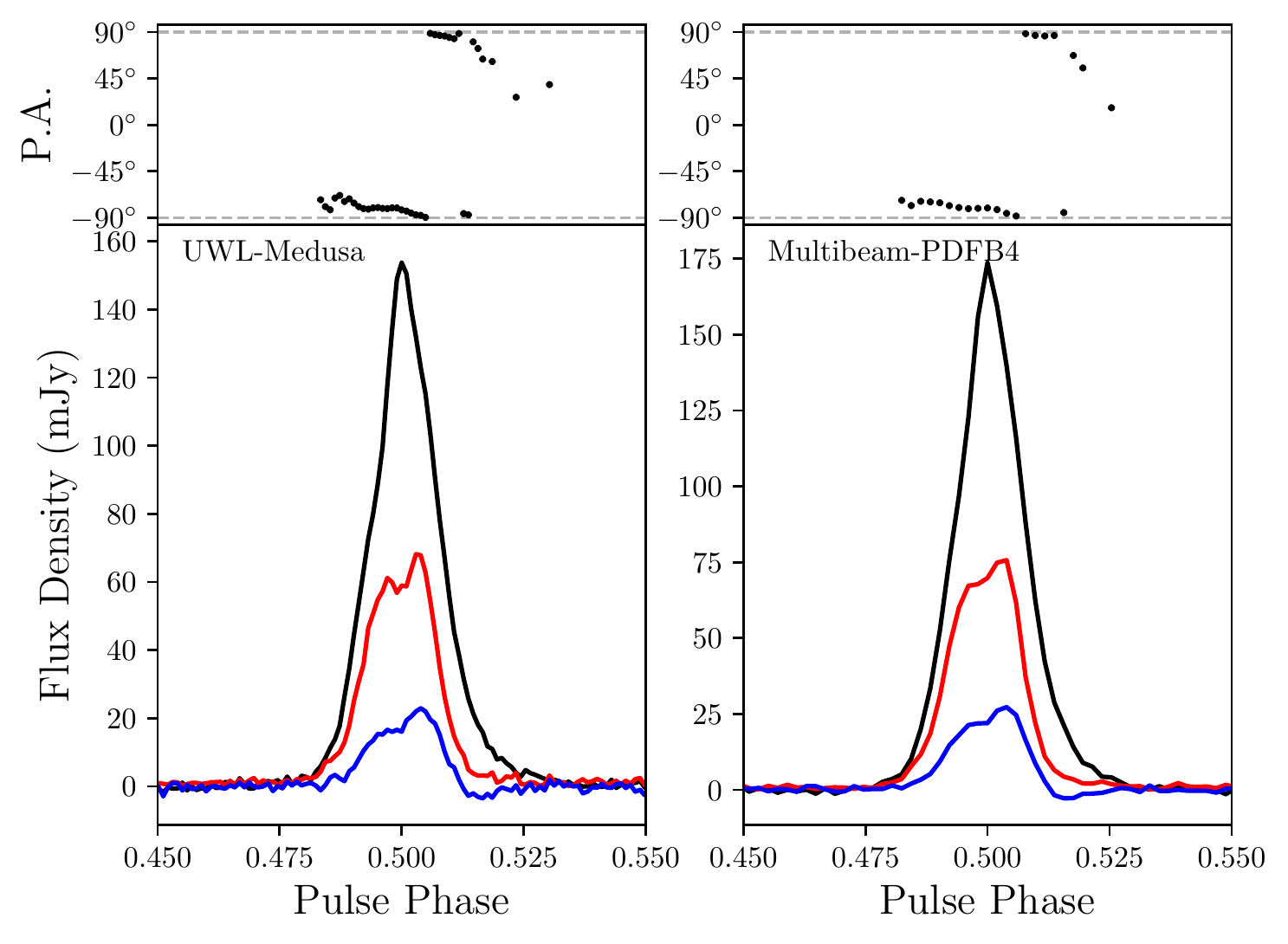}
    \caption{Comparison of the Stokes profiles for PSR~J1909$-$3744 in the 20cm band (sub-band E) observed with the Murriyang UWL receiver recorded with the Medusa backend (left) and as published by \citet{2015MNRAS.449.3223D} (right), observed with the Multibeam receiver and recorded by the PDFB4 backend. The bottom panels show the total intensity (black), linear polarisation (red) and circular polarisation (blue), with the linear polarisation angle (PA) shown in the top panels. This demonstrates the consistency between the recent UWL observations and those taken with earlier systems. Note that the \citet{2015MNRAS.449.3223D} observation is recorded with 512 phase bins as opposed to 1024 as in the UWL observation.}
    \label{fg:compare1909}
\end{figure}

\begin{figure}
    \centering
    \includegraphics[width=\linewidth,angle=0]{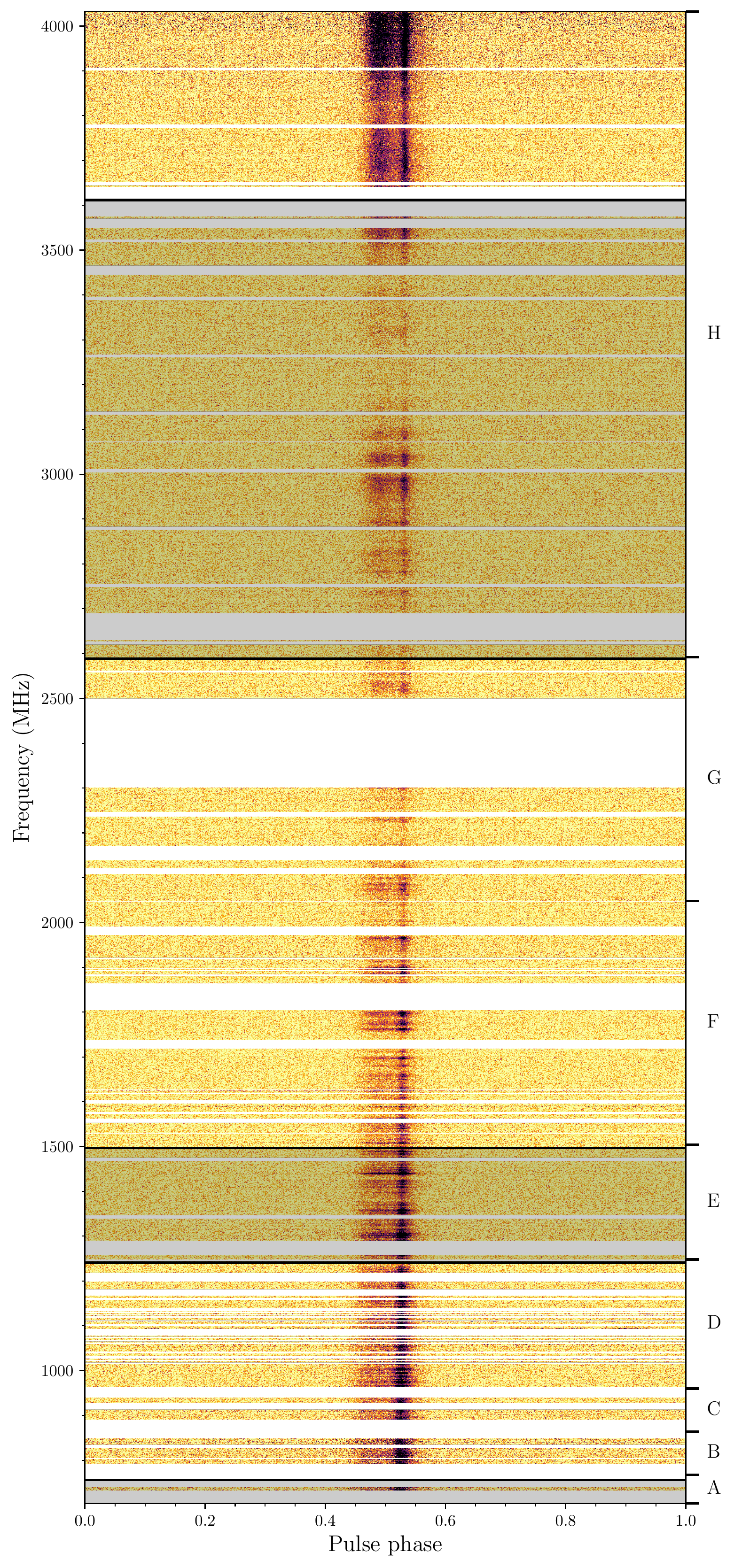}
    \caption{Normalised intensity as a function of observing frequency and pulse phase from a single wide-band observation of PSR~J1600$-$3053. The shaded regions indicate the frequency coverage of the previous receivers used for PPTA observations, highlighting the additional frequency coverage provided by the UWL. The frequency range of each UWL sub-band is indicated on the right-hand abscissa. Flux density variations across frequency are caused by interstellar scintillation. Frequency ranges that have been flagged out due to interference are left blank.}
    \label{fg:uwl1600}
\end{figure}

\begin{figure*}
    \centering
    \includegraphics[width=\linewidth]{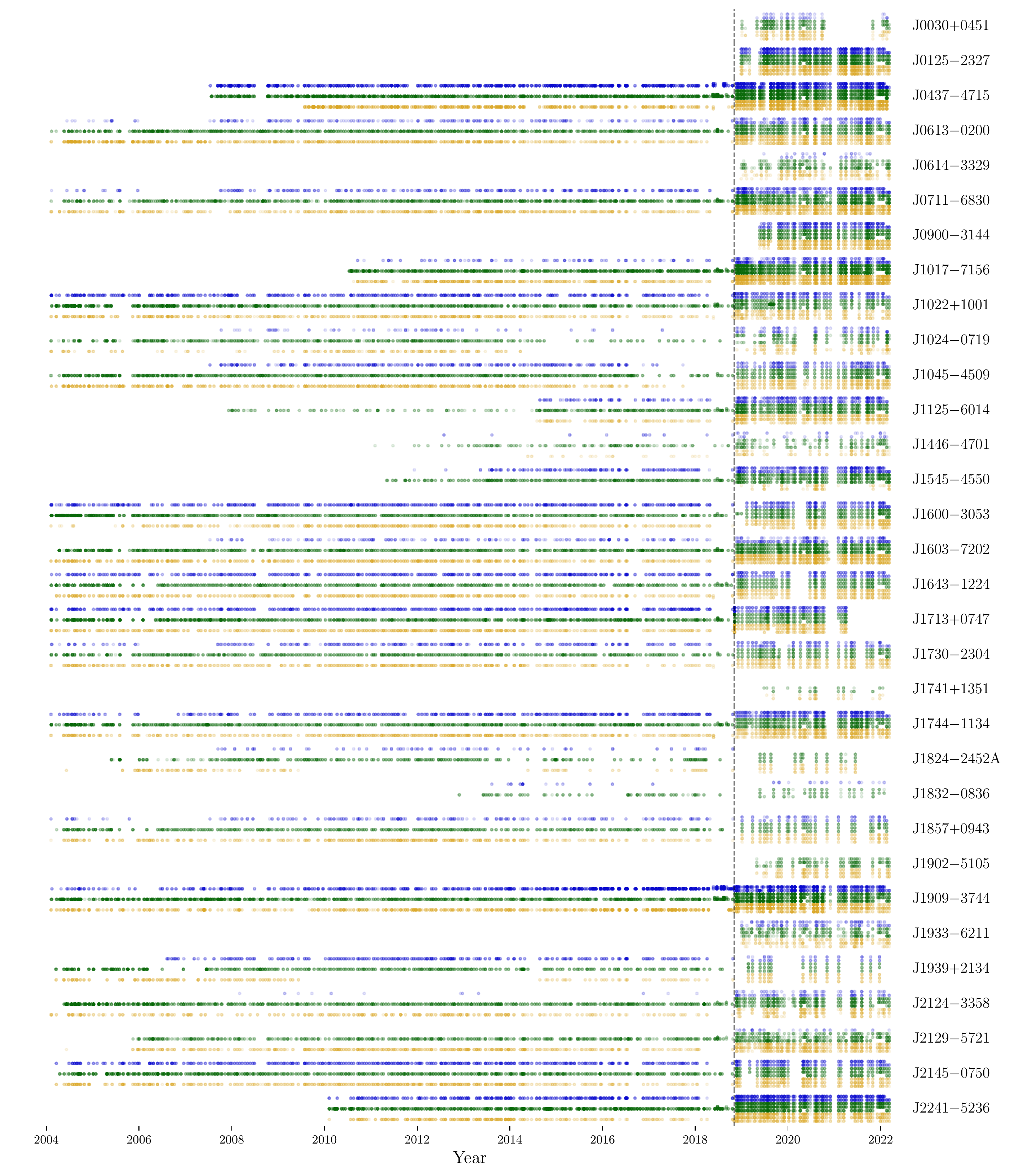}
\caption{Observing times of the PPTA DR3. The vertical dashed line indicates the beginning of UWL observations. The top, middle, and lower lines in pre-UWL observations of each pulsar represent observations in 10\,cm, 20\,cm, and 40\,cm wave-bands indicated by blue, green, and gold colours respectively. Observations taken with the UWL are split into the UWL sub-bands (see Table \ref{tb:subbands}). UWL sub-bands A, B, and C cover  the 40\,cm band, D, E, F cover the 20\,cm band, and G, H cover the 10\,cm band.}   
    \label{fg:dr3obs}
\end{figure*}

The PPTA Data Release Three (DR3) combines reprocessing of the earlier data release with new ultra-wide bandwidth receiver observations. 
In this release we have chosen to perform narrow-band pulsar timing, in which times of arrival (ToAs) are derived from sub-banded data, using frequency dependent pulse portraits.
This approach allows us to correct for the effect of DM variations while also accounting for the intrinsic frequency dependence of the pulse profiles. It also provides the most natural combination of the legacy and UWL data sets. 
 The data release also contains the calibrated pulse profiles and dynamic spectra corresponding to the most recent observations.  

The key properties of our data set are presented in Table \ref{tb:dr3}.
The first four columns of that Table list the pulsars that are included in the data release along with their pulse period, DM and orbital period. Twenty-five of the pulsars have been described in earlier PPTA papers (e.g., \citealt{2020PASA...37...20K}). We excluded one of the pulsars included previously in DR2, PSR~J1732$-$5049, because regular observations ceased in 2011 owing to its large ToA uncertainties. Compared to previous data releases, we include a further six binary pulsars (PSRs~J0125$-$2327, J0614$-$3329, J0900$-$3144, J1741$+$1351, J1902$-$5105 and J1933$-$6211) and one solitary pulsar (PSR~J0030$+$0451).
PSRs~J0900$-$3144, J1741$+$1351 and J1933$-$6211 were discovered in Murriyang surveys \citep{2006MNRAS.368..283B} and first published in 2006 (PSR~J0900$-$3144) and 2007. PSRs~J0614$-$3329 and J1902$-$5105 were discovered in surveys targeting unidentified Fermi-LAT sources by the Green Bank Telescope \citep{2011ApJ...727L..16R} and Murriyang \citep{2012ApJ...748L...2K}. PSR~J0125$-$2327 was discovered more recently, initially in the Green Bank North Celestial Cap survey \citep[GBNCC;][DeCesar et al. \textit{in prep.}]{2020ApJ...892...76M}, and later was independently discovered in the Parkes High Time Resolution Universe survey \citep[HTRU; ][]{2019MNRAS.483.3673M}. These pulsars were not previously included in PPTA observations, but were added after commissioning of the UWL receiver. This was possible largely due to the improved observing efficiency enabled by the receiver (see \S\ref{subsec:widebanddiscussion}). 

In the subsections below we first describe the UWL observations, their corresponding auxiliary data products and then how we have combined those observations with the previous data release.

\subsection{The wide-bandwidth observations and processing}\label{sec:uwl}

This is the first data release presenting data obtained with the UWL system. The UWL system resulted in a change in observing strategy. Previously in each session, every pulsar was typically observed at least twice with two different narrow-band receiver systems, to measure and correct for DM variations. With the UWL system, this strategy is no longer necessary, thanks to its wide, continuous frequency coverage.

Before observing each pulsar, we took an observation at a nearby reference position, injected with the signal from an artificial noise source. The noise source is switched on and off at a frequency of 62.5\,Hz, and is injected equally into the low noise amplifiers for the two polarisation streams. We used the observations with the artificial noise source for flux density and polarisation calibration. During each observing session, we also observed a primary flux calibrator source (usually PKS~B0407$-$658 or PKS~B1934$-$638\footnote{Previous flux density measurements were calibrated against the radio galaxy Hydra A (3C 218, PKS 0915$-$118).  However, this source was found to be resolved at the top of the UWL band, so is not suitable for flux density calibration.}, enabling the artificial noise source to be scaled in astronomical units. At less regular intervals, we observed PSR~J0437$-$4715 from rise-to-set to enable ellipticity in the polarization response of the receiver system to be modelled and corrected for during the calibration stages (this is based on an updated version of the MEM algorithm in \citealt{2004ApJS..152..129V}).

The raw data files for each observation were recorded in PSRFITS format \citep{2004PASA...21..302H} and are stored in the CSIRO Data Access Portal (DAP) \footnote{\url{data.csiro.au}} \citep{2011PASA...28..202H} under the project identifier P456. Observations typically last 3840\,s (although some observations are shorter, either by plan or due to observations being aborted for reasons such as high winds) and we typically observe each pulsar every 2--3 weeks. 

The majority of the observations were recorded with the ``Medusa'' astronomy signal processor, which coherently de-disperses the data stream and folds the data at the known pulsar period. The raw data files are relatively large and are split into three time segments for easier archiving. In our processing steps below we combine these files back to a single observation file per observation. At certain times the Medusa backend was not available and, for those observations, we reverted back to the earlier backend instruments (CASPSR and PDFB4 as described in \citealt{2020PASA...37...12H}).  We also recorded a small number of observations with these older systems (and older receivers) to measure timing offsets between the two systems. We carried out the majority of these observations (95\%) between MJD 58256 and 58752 (2018-05-18 to 2019-09-26 UT), but continued to take a small number of observations with earlier instrumentation until close to the end of the data set.

\subsubsection{Wide-band data processing}

Our data-reduction pipeline uses standard tools and methods. 
We primarily used the \textsc{pfits} \citep{2021ascl.soft04013H} and \textsc{psrchive} software packages \citep{2004PASA...21..302H} to process the UWL observations. Initially, we installed the original timing ephemerides published by \cite{2021MNRAS.507.2137R} (apart from PSR~J0437$-$4715, which was published in \citealt{2020PASA...37...20K}), using \textsc{pam} \citep{2012AR&T....9..237V}. We performed initial radio frequency interference (RFI) flagging using the \textsc{pfits\_zapUWL} routine from \textsc{pfits}, which compares the system flux density as measured using the switched noise source with the expected values, and flags any channels that show significant deviations. We applied flux density and polarisation calibration to the observations using the \textsc{pac} routines making use of the primary flux density calibration solutions and the rise-to-set observations of PSR~J0437$-$4715. We then applied a second stage of RFI-flagging using \textsc{MeerGuard}, a modified version of the \textsc{CoastGuard} RFI flagging tool \citep{2020ascl.soft03008L}. The data for each observation file are averaged in time.

The most recent data in the PPTA second data release provided observations in three observing bands corresponding to the three receivers used: 40, 20 and 10\,cm-bands. To simplify the concatenation of the earlier data with those from the new UWL system, we split the UWL data into similar bands. In total, we divided the data into the eight bands listed in Table~\ref{tb:subbands}. Bands~A, E and H are close to the original bands used in the second data release. The sub-band bandwidths increase with increasing frequency to improve the uniformity of signal-to-noise ratios (S/N), owing to the decrease in MSP flux densities with frequency \citep{1999ApJ...526..957K}. We then averaged the data in each sub-band to form four frequency channels per sub-band. For timing applications we also formed Stokes $I$ data products.

\begin{table}[ht]
\caption{UWL sub-band frequency ranges. The minimum and maximum frequencies of each sub-band are labelled $\nu_\mathrm{min}$ and $\nu_\mathrm{max}$, respectively. We also note the wave-band that corresponds to each sub-band.\label{tb:subbands}}
\centering
\begin{tabular}{ccccc}
\hline
Sub-band & $\nu_\mathrm{min}$ & $\nu_\mathrm{max}$ & Bandwidth & Wave-band \\
label    & (MHz) & (MHz) & (MHz) & \\

\hline
    A & 704 & 768 & 64 & 40\,cm \\ 
    B & 768 & 864 & 96 & 40\,cm \\ 
    C & 864 & 960 & 96 & 40\,cm \\ 
    D & 960 & 1248 & 288 & 20\,cm \\ 
    E & 1248 & 1504 & 256 & 20\,cm \\ 
    F & 1504 & 2048 & 544 & 20\,cm \\ 
    G & 2048 & 2592 & 544 & 10\,cm \\ 
    H & 2592 & 4032 & 1440 & 10\,cm \\ 
\hline
\end{tabular}
\end{table}

To demonstrate the fidelity of our observations and our sub-banding method, in Figure~\ref{fg:compare1909} we present a single observation of PSR~J1909$-$3744 taken with the UWL in sub-band E (1248--1504\,MHz), along with a previous observation recorded using the earlier Multibeam receiver with the PDFB4 backend \citep{2015MNRAS.449.3223D} across a similar frequency range. The polarisation properties are consistent between the different observing systems, validating the performance of the new observing system and our calibration process. To demonstrate the additional observing band that is now available, in Figure~\ref{fg:uwl1600} we show a wide-band profile from a single observation of PSR~J1600$-$3053 with the previous observing bands overlaid.

\subsubsection{Timing measurements from wide-band observations}

The primary data products in our data release are the pulsar ToAs, and their associated timing residuals and ephemerides. We formed pulse ToAs by cross-correlating an idealised profile template with each observation.  For most pulsars this was conducted on the total intensity (Stokes $I$). In the case of PSR~J0437$-$4715, we first formed the invariant interval\footnote{The invariant interval is defined as $S_{\text{inv}} = {I^2 -  Q^2  - U^2 - V^2}$} \citep{2000ApJ...532.1240B} of the profile prior to forming standard templates and hence ToAs. In this data release, we first followed the method presented by \cite{2014ApJ...790...93P} to form wide-band profile portraits \citep{2023ApJ...944..128C}. We created one-dimensional templates for each channel across the eight sub-bands by evaluating the wide-band portrait model at the channel centre frequency. We then measured ToAs using the Fourier-domain Monte Carlo of the \textsc{pat} routine within \textsc{psrchive}. 

We list the total data span, number of independent observations, and the total number of ToAs from UWL observations in columns 5, 6 and 7 of Table~\ref{tb:dr3}. For each of the eight UWL sub-bands we also report the median ToA uncertainty in Table~\ref{tb:uwl_obs}. As expected, there is a range of median ToA uncertainty values across different pulsars and different observing bands (and bandwidths). In the best cases (e.g., for PSR~J0437$-$4715) the median ToA uncertainties are tens--to--hundreds of nanoseconds. In the worst cases, they are a few microseconds.

We formed timing residuals initially using the timing ephemerides presented by \cite{2021MNRAS.507.2137R} and \citet{2023ApJ...944..128C}, with the TT(BIPM2020) reference timescale published by the Bureau International des Poids et Mesures (BIPM) and the Jet Propulsion Laboratory Solar System ephemeris DE436. The wide-band nature of the observations means that the ToA precision in each sub-band can vary with the scintillation state of the pulsar. In some cases, the S/N of the profile in a sub-band was low and the ToA uncertainty was poorly determined. This led to numerous outlier residuals, which were filtered with a S/N$<10$ cut off, and not included in our data release. These low S/N observations do not inform on the pulsar timing model or provide sensitivity in GW searches.

Not all the PPTA UWL observations are presented in this data set. In particular, PSR~J1713$+$0747 underwent an event possibly originating in the pulsar magnetosphere on MJD~59321 \citep[April 2021;][]{2021ATel14642....1X,2021MNRAS.507L..57S,2022arXiv221012266J} which significantly changed its pulse profile. We defer detailed analysis of this event with our wide-band observations to a later paper, and here only include observations of this pulsar until just prior to the event. 
\begin{table*}
\caption{Median ToA uncertainties for each sub-band from UWL observations.\label{tb:uwl_obs}}
\begin{tabular}{ccccccccc}
\hline
PSR & $\sigma_A$ & $\sigma_B$ & $\sigma_C$ & $\sigma_D$ & $\sigma_E$ & $\sigma_F$ & $\sigma_G$ & $\sigma_H$\\
  & ($\mu$s) & ($\mu$s) & ($\mu$s) & ($\mu$s) & ($\mu$s) & ($\mu$s) & ($\mu$s) & ($\mu$s)  \\
  \hline
J0030$+$0451  & 3.81  & 3.13 & 3.76 & 2.74 & 3.25 & 3.34 & 4.43 & 4.12 \\
J0125$-$2327  & 2.98  & 1.97 & 2.06 & 0.90 & 0.79 & 0.55 & 0.72 & 0.68 \\
J0437$-$4715  & 0.31  & 0.22 & 0.14 & 0.08 & 0.06 & 0.07 & 0.11 & 0.14 \\
J0613$-$0200  & 0.81  & 0.60 & 0.96 & 0.77 & 1.26 & 1.50 & 2.56 & 2.51 \\
J0614$-$3329  & 3.16  & 2.83 & 2.73 & 2.25 & 2.47 & 2.39 & 1.92 & 2.43 \\
J0711$-$6830  & 4.67  & 3.78 & 3.80 & 2.75 & 2.99 & 3.34 & 3.89 & 4.53 \\
J0900$-$3144  & 10.11 & 4.15 & 4.98 & 2.31 & 2.89 & 3.09 & 4.65 & 5.02 \\
J1017$-$7156  & 1.20  & 0.67 & 0.78 & 0.45 & 0.54 & 0.70 & 1.34 & 1.65 \\
J1022$+$1001  & 3.00  & 3.23 & 3.34 & 1.87 & 1.82 & 1.98 & 2.33 & 2.03 \\
J1024$-$0719  & 2.64  & 2.80 & 2.05 & 1.63 & 1.80 & 2.46 & 3.68 & 4.11 \\
J1045$-$4509  & 5.87  & 3.71 & 5.16 & 2.97 & 4.19 & 5.18 & 9.30 & 9.74 \\
J1125$-$6014  & 1.33  & 0.99 & 1.07 & 0.60 & 0.59 & 0.55 & 0.72 & 0.68 \\
J1446$-$4701  & 1.77  & 1.39 & 1.13 & 1.18 & 1.32 & 1.29 & 1.34 & 1.68 \\
J1545$-$4550  & ---   & 2.60 & 2.44 & 1.55 & 1.18 & 0.85 & 1.09 & 1.07 \\
J1600$-$3053  & 2.20  & 1.26 & 1.30 & 0.57 & 0.51 & 0.50 & 0.83 & 1.02 \\
J1603$-$7202  & 4.47  & 2.50 & 3.24 & 1.57 & 2.05 & 2.58 & 4.20 & 4.97 \\
J1643$-$1224  & 3.32  & 1.74 & 2.33 & 1.22 & 1.37 & 1.54 & 2.21 & 2.52 \\
J1713$+$0747  & 2.12  & 0.97 & 0.93 & 0.36 & 0.31 & 0.24 & 0.34 & 0.46 \\
J1730$-$2304  & 3.78  & 2.30 & 3.19 & 1.41 & 2.10 & 2.13 & 3.54 & 2.50 \\
J1741$+$1351  & 2.00  & 1.57 & 1.61 & 0.89 & 1.33 & 1.43 & 1.56 & 1.80 \\
J1744$-$1134  & 1.28  & 0.76 & 0.95 & 0.47 & 0.57 & 0.64 & 0.83 & 0.86 \\
J1824$-$2452A & 1.65  & 0.98 & 0.89 & 0.48 & 0.51 & 0.59 & 0.84 & 0.92 \\
J1832$-$0836  & ---   & 1.30 & 1.43 & 1.07 & 1.00 & 0.87 & 0.89 & 0.80 \\
J1857$+$0943  & 5.32  & 3.55 & 3.75 & 2.00 & 2.06 & 1.46 & 2.87 & 2.33 \\
J1902$-$5105  & 1.74  & 1.15 & 1.56 & 1.35 & 2.08 & 2.27 & --- & --- \\
J1909$-$3744  & 0.33  & 0.26 & 0.28 & 0.18 & 0.19 & 0.15 & 0.17 & 0.15 \\
J1933$-$6211  & 2.86  & 3.06 & 3.12 & 2.71 & 3.12 & 2.85 & 3.23 & 3.38 \\
J1939$+$2134  & 0.11  & 0.06 & 0.08 & 0.05 & 0.06 & 0.07 & 0.14 & 0.16 \\
J2124$-$3358  & 3.94  & 3.53 & 4.09 & 3.32 & 3.86 & 4.53 & 6.74 & 7.03 \\
J2129$-$5721  & 2.66  & 2.19 & 2.74 & 2.27 & 2.86 & 3.03 & 3.63 & 3.08 \\
J2145$-$0750  & 2.55  & 2.35 & 2.79 & 1.25 & 1.67 & 1.41 & 1.98 & 2.67 \\
J2241$-$5236  & 0.43  & 0.33 & 0.38 & 0.28 & 0.36 & 0.47 & 0.87 & 0.86 \\
\end{tabular}
\end{table*}

\subsubsection{Auxiliary wide-bandwidth data products}

In addition to the data products used for pulsar timing, we provide products that are useful for other scientific purposes.
For each observation, we formed dynamic spectra with the full frequency (1\,MHz) and time resolution. The time resolution for observations before MJD 57802 is 30\,s, and after this, 10\,s. We used the \textsc{psrflux} routine with the wide-band template as the standard profile to form the dynamic spectra. 

As well as producing pulsar fold-mode data products, the Murriyang UWL observing system allows simultaneous observations either in pulsar search mode, or with high frequency-resolution and low time-resolution mode (a spectral line mode). The high-frequency-resolution data sets provide an easy way to inspect the RFI environment during our observations or other science goals \citep[e.g., the search for dark matter axions; ][]{2017ApJ...845L...4K}. Only a relatively small subset of PPTA observations are recorded with spectral line mode. The raw data for the high frequency resolution observations are available from the Australia Telescope Online Archive\footnote{\url{https://atoa.atnf.csiro.au/}}, and have been recorded across the entire band with 30\,Hz frequency resolution and 10-second spectral dump times. The data are in SDHDF format (Toomey et al., \textit{submitted}). In our data collection we provide the spectral line data averaged in time, but with the native frequency resolution.

\subsubsection{Instrumental improvements and systematic effects}
Various improvements have been made to the UWL system since its installation. For instance, on 2022-03-17 (MJD~59655) notch filters, to reduce persistent RFI, and a new noise injection system were installed. We have not completely characterised the new noise source yet and hence the last observation in the third data release described here is before these updates were made. There have also been failures in the system. The timing and synchronisation systems have required multiple restarts. While we have carefully searched for and corrected timing offsets (see $\S$ \ref{subsubsec:jumps}), these restarts may have caused subtle timing offsets that are either marginally or un-detectable within single-pulsar timing residuals. Additionally, de-synchronisation of polarisation channels led to no signal being recorded in cross-polarisations for a period of several weeks in early 2021. There are also time ranges where attenuation levels were not optimally set, resulting in artefacts in the pulse profiles. Further, until the notch filters were introduced, strong RFI was affecting the quality of the switched noise source and hence slightly degraded our calibration fidelity. 

\subsubsection{Timing offsets}\label{subsubsec:jumps}
Timing offsets of up to $\sim 1\,\mu$s can be introduced into pulsar ToAs through changes to the timing delays along the signal processing chain. These could be, for example, changes in the observing backend, resets of the timing, synchronisation, and signal digitisation systems, hardware modifications, among others. In the course of producing our data sets, we have made concerted efforts to search for and measure timing offsets \citep{2020PASA...37...20K}, which are corrected using \textsc{tempo2} JUMP parameters in the timing model parameter file.

Our search for timing offsets within the new UWL data was initially performed through manual inspection of the timing residuals while producing the initial timing models and single-pulsar noise models. For changes in observing systems (e.g., in the transition to the Medusa system from the older PDFB4 and CASPSR systems), data recorded simultaneously with both backends were used to constrain the timing offset between them. In the course of our manual processing, we identified an additional timing offset at MJD~59200 with an amplitude of approximately 120\,ns. The cause of this offset likely to be a digitiser resynchronisation that occurred a few days prior to the observation, after a data transfer issue internal to the Medusa backend was identified. We correct this timing offset using a floating \textsc{tempo2} JUMP parameter applying to MJD 59200. We searched for additional timing offsets using a parameter estimation approach in our single-pulsar noise analysis, described in our noise modelling companion paper \citep{PPTA-DR3_noise}. The reference system for the JUMP parameters is 10-cm PDFB4, as in DR2.

\subsection{Combining with the PPTA second data release}

We have combined our new UWL ToAs with those published as our second data release (DR2) to form our third data release (DR3). Details of the second data release were provided by \citet{2020PASA...37...20K}, which included calibrated and RFI-excised pulse profiles, each with 32 frequency channels\footnote{The data release 2 collection is available from \url{https://doi.org/10.25919/5db90a8bdeb59}.}. We have re-processed these pulse profiles, to produce new ToAs, which are compatible with the UWL data.

The steps involved in re-processing of the PPTA-DR2 pulse profiles were as follows:
\begin{itemize}
    \item \textsc{MeerGuard} software package was applied to the previously calibrated and RFI-excised pulse profiles, to improve the rejection of RFI.
    \item The latest timing ephemeris and the DM derived from UWL data \citep[described in][]{2021MNRAS.507.2137R} were installed.
    \item Each observation was reduced to four frequency channels. 
    \item A frequency-dependent template was generated from the analytical wide-band portraits derived from the UWL data.
    \item The four channels for each observation were timed with the corresponding channel from the template, using the Fourier-domain Monte Carlo method of \textsc{pat} within \textsc{psrchive}.
\end{itemize}
 We also processed, in a similar fashion, additional data that were recorded with the PDFB4 and CASPSR backend systems following the end of the PPTA-DR2. Where data were recorded by more than one backend simultaneously (e.g. with PDFB4 and Medusa), we only preserve ToAs recorded with the older system within the relevant, overlapping frequency range (e.g., we would discard a ToA recorded from sub-band E with Medusa in favour of a ToA from PDFB4 taken from the same observation). This avoided duplication of ToAs and allowed us to constrain the timing offset between Medusa and older backend systems.

The ToAs were combined to produce our timing data set for each pulsar. The  data span, total number of observations and number of individual ToAs are listed in Table~\ref{tb:dr3}. Note that we have excised $\sim 3.4$\,yr-worth of ToAs from the beginning of the data set for J0437$-$4715. These ToAs were recorded with early systems (CPSR2, PDFB1, and WBCORR), and exhibited inconsistent frequency-dependent delays that were attributed to frequency- and phase-dependent sensitivity of these older systems \citep{2020PASA...37...20K}.

\begin{figure*}
\includegraphics[width=\textwidth]{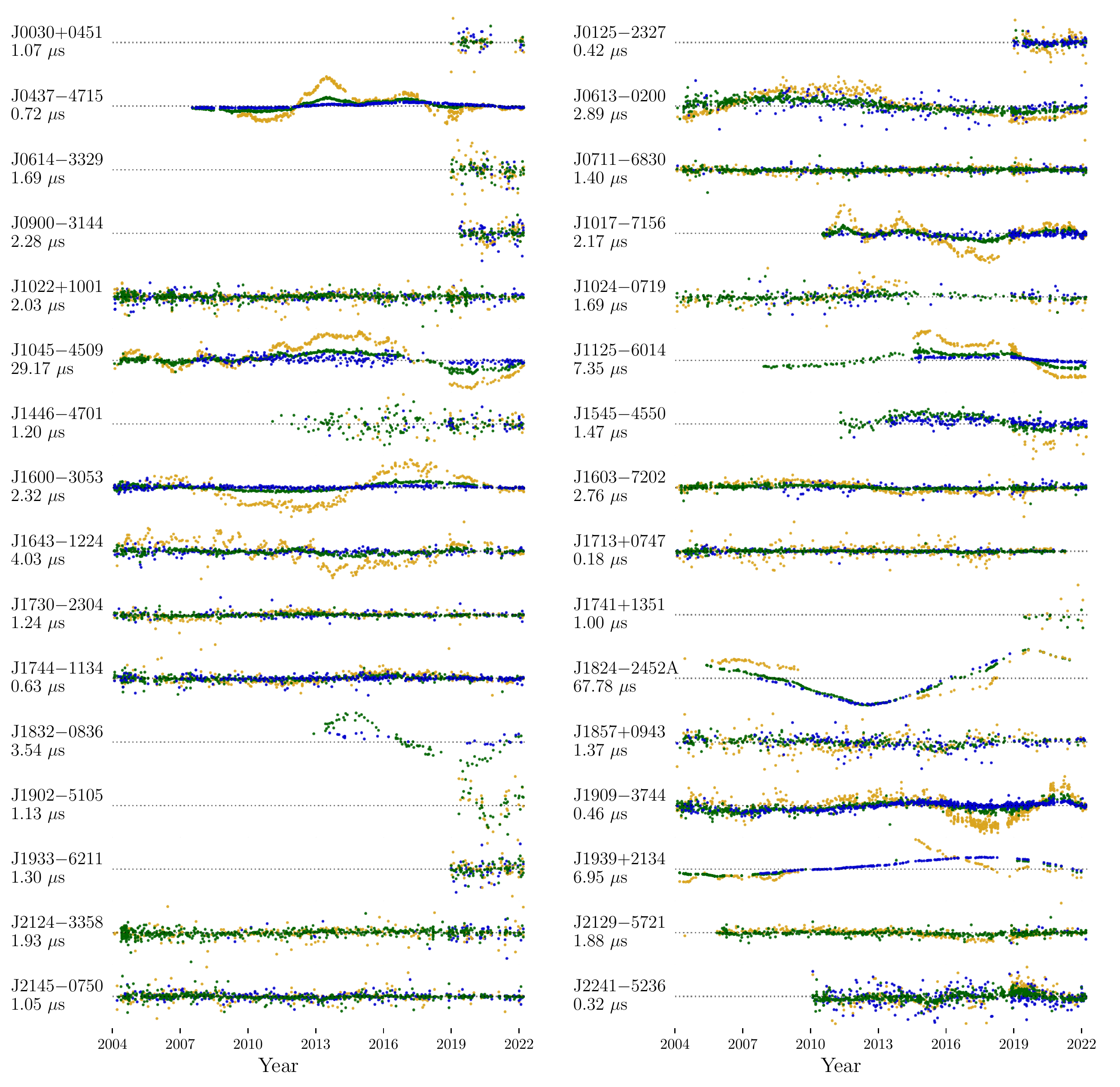}
\caption{Band-averaged timing residuals for the PPTA DR3 formed under the assumption of our detailed single-pulsar noise models, as described in the companion PPTA-DR3 noise analysis paper \citep{PPTA-DR3_noise}. The colours indicate frequency bands as in Figure \ref{fg:dr3obs}. We show the weighted rms across all frequencies beneath each pulsar label. Note that we have not subtracted any chromatic or achromatic noise processes from these residuals.}
\label{fg:residuals}
\end{figure*}

\begin{figure*}
\includegraphics[width=\textwidth]{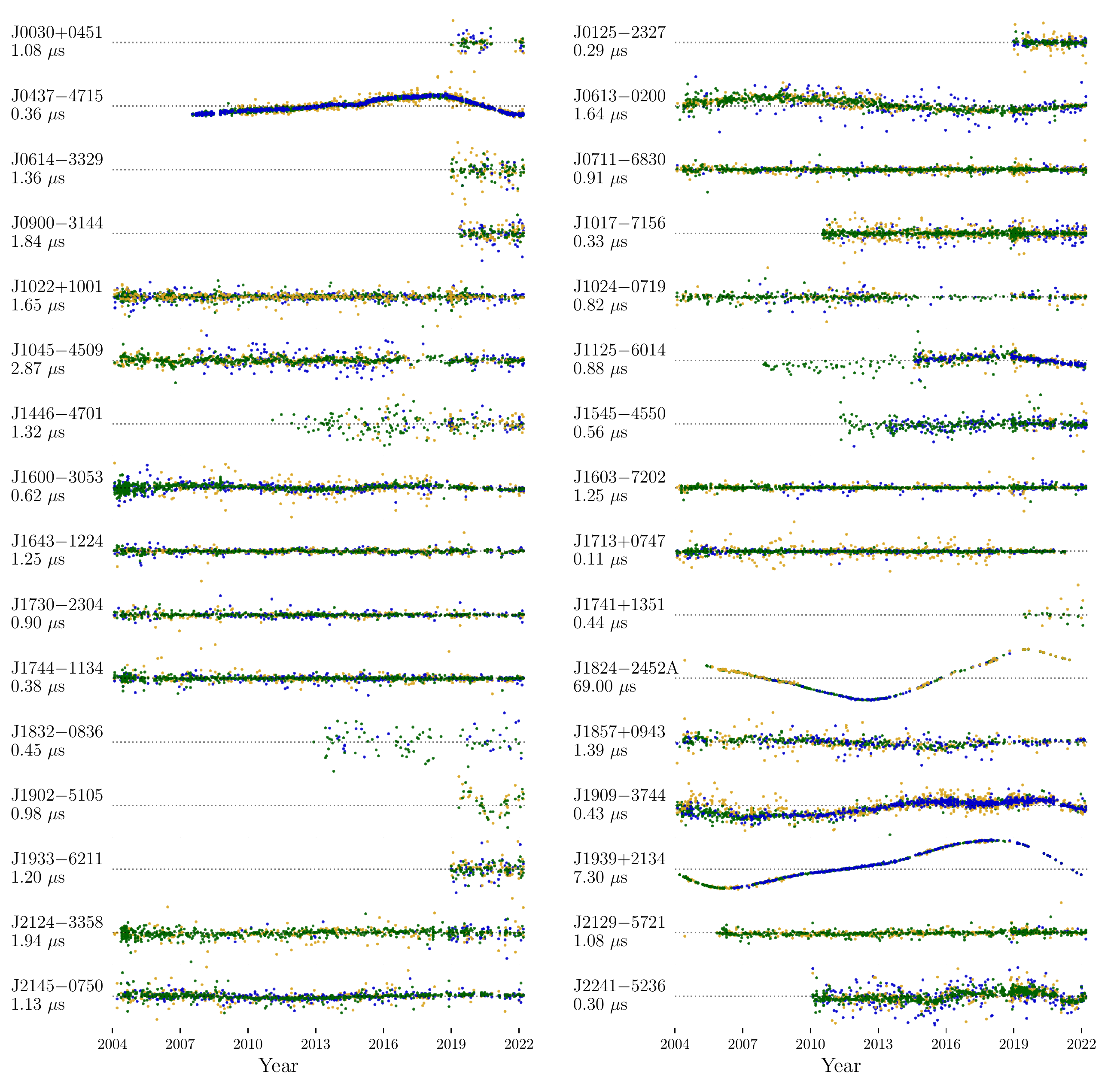}
\caption{Same as in Fig. \ref{fg:residuals}, but with all frequency-dependent and system-dependent noise terms subtracted, leaving only achromatic red noise present in the residuals.}
\label{fg:achrom_residuals}
\end{figure*}

We produced an initial timing model fit for all pulsars using the ephemerides from \citet{2021MNRAS.507.2137R}, \citet{2023ApJ...944..128C}, and Mandow et al. (\textit{in prep.}). We provide the derived ephemerides for each pulsar in the data release repository. Our final timing models presented in this data release are physically consistent, and sufficiently whiten the timing residuals for the purpose of noise modelling. However, we defer a detailed analysis of the astrophysical implications of our timing models to a future paper. We present our detailed single-pulsar noise models in a companion paper \citep{PPTA-DR3_noise}. To provide a representation of our timing residuals, in Figure~\ref{fg:residuals} we show timing residuals averaged into 40\,cm, 20\,cm, and 10\,cm wavelength bands for each pulsar, produced assuming our detailed single-pulsar noise models, but without subtracting time-domain realisations of any of the noise processes described in \citet{PPTA-DR3_noise}. In Figure~\ref{fg:achrom_residuals}, we show the same residuals after subtracting time-domain realisations of frequency- and system-dependent noise processes, leaving behind only achromatic red noise. These figures illustrate the data spans, sampling, and the presence of low-frequency chromatic and achromatic noise present in the pulsars. 

\section{Discussion}\label{sec:discussion}

\subsection{The wide-bandwidth receiver}
\label{subsec:widebanddiscussion}
The UWL receiver system provides higher quality (in terms of bandwidth, system temperature and hence rms timing residuals) observations than earlier observing systems. Additionally, an important practical outcome of the increased bandwidth provided by the UWL is the ability to simultaneously cover the same frequency bands as pre-UWL observations in a single observation, without the need to switch between receivers. This has substantially improved our observing efficiency, enabling us to add several MSPs into our observing program without increasing our total time request. A detailed analysis of the profiles and timing parameters for the newly-added MSPs will be presented elsewhere, but we include arrival times and basic timing model parameters for a subset of the newly added pulsars as part of this paper.

\begin{figure*}
    \centering
    \includegraphics[angle=-90,width=17cm]{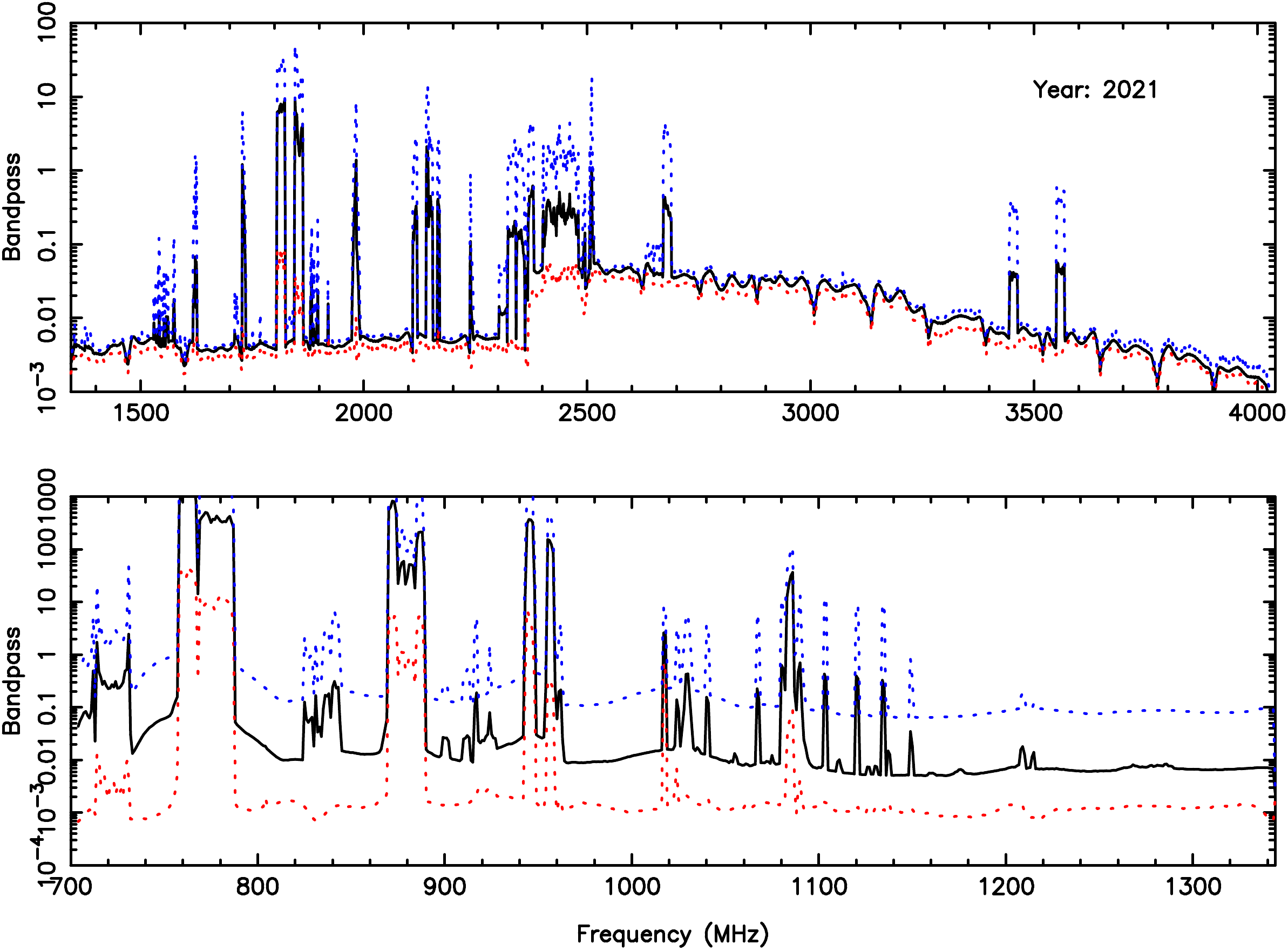}
    \caption{Bandpass of UWL PPTA observations during the year 2021. The black line indicates the mean bandpass, the minimum bandpass signal is in red and the blue line indicates the highest signal detected. The sudden increase in power at 1344\,MHz and 2368\,MHz is caused by changing attenuation levels between separate digitisers; see \citealt{2020PASA...37...12H} for details.}
    \label{fg:uwl_rfi}
\end{figure*}

The primary challenge with wide-bandwidth systems is the RFI environment. Wide-band systems need to be able to perform well in the presence of any strong interference source in the band. In Figure~\ref{fg:uwl_rfi} we present our typical bandpass (black) overlaid on the minimum (red) and maximum (blue) signal strength detected during the year 2021. These bandpass measurements have been obtained from the noise-source observations of the PPTA prior to the observation of PSR~J0437$-$4715. The bandpass is divided into the lowest part of the band, from 704 to 1344\,MHz, where the RFI is strongest and the remainder of the band (from 1344 to 4032\,MHz). Much of the RFI is always present, however, mobile handset transmissions around 700\,MHz and WiFi/Bluetooth signals around 2.4\,GHz are more dominant during the day. 

To determine the typical fraction of 1\,MHz frequency channels that were flagged, we investigated our processed observations of PSR~J1909$-$3744. On average, 26\% of the channels were removed, corresponding to $\sim$850\,MHz. Approximately 200\,MHz is typically removed around 2.4\,GHz, due to WiFi/Bluetooth and Australian National Broadband Network transmissions. The majority of other interference sources are in the low-frequency part of the band and related to mobile transmission towers and handsets. We also flag $\sim$6\,MHz at each of the 26 sub-band boundaries due to bandpass roll off (see \citealt{2020PASA...37...12H}).

\subsection{Comparison with earlier work}
\label{sec:comparison}

The time-span covered by our DR3 data set (MJD 53040--59640) is approximately three years longer than DR2. For the majority of our pulsars, our data spans between 10 and 18 years (without the inclusion of the legacy data sets). Our data spans are therefore now significantly longer than the orbital period of Jupiter (12 years) and the Solar cycle ($\sim 11$ years), but the high quality data sets only represent approximately one half of the orbit of Saturn. This will enable an improved ability to identify the potential impact of errors in the Jovian orbital parameters \citep[e.g.][]{2020ApJ...893..112V}, or the effects of the solar wind \citep{2022ApJ...929...39H,2021A&A...647A..84T}, with other red noise processes such as pulsar spin noise \citep{2010ApJ...725.1607S}, variations in the interstellar electron density \citep{2016ApJ...817...16C}, or a GW signal.

The UWL provides demonstrated improvement in sensitivity to the legacy systems. 
To compare the earlier instruments with the UWL, we compared the timing measurements from the last year of DR2 with the last year of timing measurements from UWL observations in the same observing bands. The rms timing residuals (after fitting only for the pulse frequency, its first derivative and, if present, the Keplerian binary parameters) are listed in Table~\ref{tb:compareUWL_DR2}. For all pulsars listed, the rms timing residual for the highest frequency ToAs (band H) are reduced with the UWL. This is expected as that band is not significantly affected by RFI and the UWL receiver has lower system noise in that band than our previous receivers. In the sub-band (E) closest to the legacy 20\,cm system, similar timing residuals are obtained with both systems. The lowest frequency band (band A) has worsened. This is not unexpected as the RFI continues to worsen in that band, due to the introduction of new mobile phone systems. 
We note that these results are only indicative as we had significantly more observations in the year prior to the UWL, than in the more recent data span. 

As shown in Figure~\ref{fg:uwl1600}, the UWL provides instantaneous frequency coverage that spans previously-used frequency bands, along with bands that were not previously covered. Previously-covered frequency bands spanned approximately 692 -- 756\,MHz, 1241 -- 1497\,MHz, and 2588 -- 3612\,MHz (referred to as the 40/50\,cm, 20\,cm and 10\,cm bands respectively), which could be observed by instruments such as the 10/40\,cm dual-band receiver, the 20-cm Multibeam receiver, and H-OH receiver. The bandwidth provided by the UWL exceeds the total bandwidth offered by the old receiver fleet by a factor of $\sim 2.5$. This improves the sensitivity of the PPTA measurements, and improves observing efficiency by removing the need to switch receivers for multi-frequency coverage.

Compared with the DR2, the DR3 parameter files now include additional frequency-dependent (``FD'') parameters than previously. These parameters were required in the timing solutions to fit excess time-stationary frequency-dependent structure in the residuals. The FD-parameters \citep{2015ApJ...813...65N} are the coefficients of a polynomial function used to represent the pulse profile evolution as a function of frequency. The majority of our parameter files now contain three FD parameters, with PSR~J0437$-$4715 requiring six (previously the DR2 parameter file for this pulsar included four). \citet{2023ApJ...944..128C} describes the frequency evolution of these pulsars in detail.  

We believe the primary cause of the frequency-dependent timing residuals is subtle imperfections in the wide-band pulse portraits. Less likely to be the primary cause is intra-channel pulse profile frequency evolution -- the portraits were constructed from observations reduced to 8\,MHz frequency resolution (corresponding to 416 channels across the UWL bandwidth), which is lower than the channel bandwidth used to form our ToAs. Furthermore, we constructed portraits from observations with full 1\,MHz frequency resolution (3328 channels), and found they were consistent with the 416-channel portraits. Nonetheless, improvements in both the wide-band pulse portraits, and to a lesser extent, more careful choice of ToA channel width, may help to reduce these systematics in future analysis. We refer the reader to Section 4.3 of \citealp{2023ApJ...944..128C} for more detailed discussion on this issue.

During the course of our noise modelling, we found that additional amendments to the timing models were necessary; namely, additional ``FB'' (binary frequency) parameters for PSR~J2241$-$5236, and Shapiro delay parameters for PSRs~J0614$-$3329 and J1902$-$5105. For more detail, see \citet{PPTA-DR3_noise}.

We have chosen to present a sub-band timing method as it provides the most practical method for combining legacy and broad band observations
\citet{2023ApJ...944..128C} applied wide-band timing methods to the PPTA UWL observations, presented pulse portraits, initial timing models and timing residuals, and discussed the profile evolution of the pulsars. The wide-band portraits presented in that work were used here in ToA determination for both the UWL and the re-analysis of the DR2 data sets.

\begin{table}
    \caption{Comparison of 1\,yr of observations pre- and post- the UWL receiver system.\label{tb:compareUWL_DR2}}
\begin{tabular}{ccccc}
\hline
    PSR  & Sub-band  & $\sigma^{\rm dr2}_{\rm 1yr}$  & $\sigma^{\rm UWL}_{\rm 1yr}$  \\
         &  & ($\mu s$) & ($\mu s$)  \\
    \hline
        J1713$+$0747 & A & 0.743 & 1.172 \\
         & E & 0.109 & 0.144 \\
         & H & 0.230 & 0.215 \\ \\
        J1744$-$1134 & A & 0.534 & 0.704 \\
         & E & 0.311 & 0.322 \\
         & H & 0.762 & 0.581 \\ \\
        J1909$-$3744 &  A & 0.258 & 0.310 \\ 
         &  E & 0.127 & 0.124 \\
         &  H & 0.112 & 0.103 \\
         \hline
\end{tabular}
\end{table}

\subsection{Including the legacy data}

Murriyang observations exist for approximately 20 of the PPTA sample of pulsars for up to 11 years prior to the start of the PPTA project \citep{2008ApJ...679..675V, 2009MNRAS.400..951V}. The extended data sets can be obtained from the first data release\footnote{\url{https://doi.org/10.4225/08/534CC21379C12}} \citep[][]{2013PASA...30...17M}. In column 5 of Table~\ref{tb:dr3}, the total data span that would be achieved through the addition of this legacy data set for the relevant pulsars. We note that the frequency coverage is less complete, and timing precision is significantly worse in the legacy data compared with the most recent observations. With current analysis methods the inclusion of this earlier data is unlikely to improve the sensitivity of the data set in terms of GWB detection. For this reason we do not include it as part of the DR3. 

\section{Summary and Conclusions}

We have described the PPTA third data release, which is the first PTA data release to include observations from a receiver system with such a wide fractional bandwidth ($\Delta f / f_0 \sim 4.7$). Such wide bandwidths will become more common in the near future as similar instruments will soon be commissioned at the 100-m Effelsberg and 110-m Green Bank radio telescopes (e.g., \citealt{2020AAS...23517517B}). 

The Murriyang UWL receiver system will  continue to be upgraded which will improve the fidelity of observations. New and improved RFI-mitigation methods, along with the commissioning of an oversampled filterbank system, will significantly improve the sensitivity of the UWL observations. The Murriyang receiver suite is currently being expanded with the installation of a cyrogenically-cooled phased-array-feed. This system will operate in a frequency band between 700 and 2000\,MHz. Although it is primarily a survey instrument, it will have pulsar timing modes allowing multiple pulsars to be observed simultaneously with a lower system temperature and greater aperture efficiency than available with the UWL.

The primary goal of producing this data release is to enable searches for the GWB. A detailed description of our noise models, and our GW analysis \citep{PPTA-DR3_gwb}, will be presented alongside corresponding papers from the CPTA, EPTA, InPTA, and NANOGrav. The EPTA, InPTA, MPTA, NANOGrav, and PPTA data sets will be combined to form the third IPTA data release, which will be the most sensitive pulsar-based data set to be used in the search for the GWs. We expect that our data release (and the IPTA data release) will also be used for numerous scientific goals including searches for individual GW sources \citep[e.g.,][]{2023arXiv230103608A}, searching for irregularities in terrestrial time standards \citep[e.g.,][]{2023MNRAS.519.3976M}, and searching for currently unknown objects in our Solar System \citep{2019MNRAS.489.5573G}.

\section*{Data Availability}


The Parkes Pulsar Timing Array Data Release 3 is available in two collections on the CSIRO Data Access Portal under ``Parkes Pulsar Timing Array Third Data Release'' (DOIs: \url{https://doi.org/10.25919/w0nw-jt05} and \url{https://doi.org/10.25919/23wj-1d69}). In this data collection we provide data products relating to the UWL observations, the re-processing of PPTA DR2 and the final DR3 data collection. 
For each pulsar we provide the calibrated profiles for each UWL observation, along with dynamic spectra. We provide the wide-band templates and corresponding pulse arrival times. We also provide timing model parameter files (\texttt{.par} files).

\begin{acknowledgement}
The Parkes radio telescope (Murriyang) is part of the Australia Telescope National Facility (https://ror.org/05qajvd42) which is funded by the Australian Government for operation as a National Facility managed by CSIRO. We acknowledge the Wiradjuri People as the traditional owners of the Observatory site. We acknowledge the Wallumedegal People of the Darug Nation and the 
Wurundjeri People of the Kulin Nation as the traditional owners of the land where this work was carried out. We thank members of the IPTA for their helpful comments on this work. This paper includes archived data obtained through the Parkes Pulsar Data archive on the CSIRO Data Access Portal (\url{https://data.csiro.au}). We thank the CSIRO Information Management and Technology High Performance Computing group for access and support with the petrichor cluster. Parts of this work were performed on the OzSTAR national facility at Swinburne University of Technology. The OzSTAR program receives funding in part from the Astronomy National Collaborative Research Infrastructure Strategy (NCRIS) allocation provided by the Australian Government. Part of this research was undertaken as part of the Australian Research Council (ARC) Centre of Excellence for Gravitational Wave Discovery (CE170100004). RMS acknowledges support through ARC Future Fellowship FT190100155. ZCC is supported by the National Natural Science Foundation of China (Grant No.~12247176 and No.~12247112) and the China Postdoctoral Science Foundation Fellowship No.~2022M710429. S.D. is the recipient of an Australian Research Council Discovery Early Career Award (DE210101738) funded by the Australian Government. L.Z. is supported by ACAMAR Postdoctoral Fellowship and the National Natural Science Foundation of China (Grant No. 12103069). Work at NRL is supported by NASA.
\end{acknowledgement}



\end{document}